\documentclass[pra,twocolumn]{revtex4-1}
\usepackage[usenames,dvipsnames]{color} 
\usepackage{graphicx}
\definecolor{LinkColor}{rgb}{0,0,.5}
\usepackage[colorlinks=true,linktoc=Red,linkcolor=Red,citecolor=LinkColor,urlcolor=LinkColor]{hyperref}
\renewcommand{\emph}{\textit}
\newcommand\Id{\leavevmode\hbox{\small1\normalsize\kern-.33em1}}

\newcommand{\ham}{{\mathcal{H}}}
\newcommand{\tr}[1]{\mathrm{\footnotesize Tr}\left\{{#1}\right\}}
\newcommand{\Tr}[1]{\mathrm{\footnotesize Tr}\left\{{#1}\right\}}

\DeclareMathSymbol{\theta}{\mathalpha}{letters}{"23}
\DeclareMathSymbol{\phi}{\mathalpha}{letters}{"27}

\newcommand{\sz}{\sigma_z}
\newcommand{\sx}{\sigma_x}
\newcommand{\sy}{\sigma_y}
\renewcommand{\sp}{\sigma_+}
\newcommand{\sm}{\sigma_-}

\begin{document}

\title{Decay of spin coherences in one-dimensional spin systems}
\author{G. Kaur}
\affiliation{Department of Nuclear Science and Engineering and Research
Laboratory of Electronics,
Massachusetts Institute of Technology, Cambridge, MA, USA}
\author{A. Ajoy}
\affiliation{Department of Nuclear Science and Engineering and Research
Laboratory of Electronics,
Massachusetts Institute of Technology, Cambridge, MA, USA}
\author{P. Cappellaro}
\email{pcappell@mit.edu}
\affiliation{Department of Nuclear Science and Engineering and Research
Laboratory of Electronics,
Massachusetts Institute of Technology, Cambridge, MA, USA}

\begin{abstract}
Strategies to protect multi-qubit states against decoherence are difficult to formulate because of their complex many-body dynamics.
A better knowledge of the decay dynamics would help in the construction of decoupling control schemes.
Here we use solid-state nuclear magnetic resonance techniques  to experimentally investigate the decay of coherent multi-spin states in linear
spin chains. Leveraging on the quasi-one-dimensional geometry of  Fluorapatite crystal spin systems, we can gain a deeper insight on the multi-spin states created by the coherent evolution, and their subsequent decay, than it is possible in 3D systems. 
We are then able to  formulate an analytical model that captures the key features of the decay. 
We can thus compare the decoherence behavior for different initial states of the spin chain and link their decay rate to the state characteristics, in particular their coherence and long-range correlation among spins. Our experimental and theoretical study shows that the spin chains undergo a rich dynamics, with a slower decay rate than for the 3D case, and thus might be more amenable to decoupling techniques.
\end{abstract}
\pacs {03.67.Hk, 03.67.Lx, 75.10.Pq, 76.90.+d}
\maketitle

%%%%%%%%%%%%%%%%%%%%%%%%%%%%%%%%%%%%%%%%%%%%%%%%%%%%%%%%%%%%%%%%%%%%

\section{Introduction}
Large quantum systems hold the promise to deliver improvements  in computation and in metrology,  by exploiting entangled or squeezed states.
Unfortunately these quantum systems are  usually very fragile and plagued by problems of decoherence \cite{Zurek91} -- as  they undergo irreversible decay \cite{Fischer09b} due to  interaction with their environment.
The decay of single qubits under the effect of various types of environments have been extensively studied and  control sequences that  could mitigate  decoherence effects have been introduced (e.g. dynamical decoupling techniques~\cite{Viola99b,Biercuk09,Ryan10,Alvarez10b}). However,  large scale quantum information processing (QIP) systems will require the preparation and control of \textit{multi-qubit} states.
These states are harder to control and to model analytically because of their complex many-body dynamics. While some recent works have looked at their decoherence and control schemes via dynamical-decoupling~\cite{Mahesh11},  the decay was usually assumed to be induced by an uncorrelated bath, acting independently on each qubit, although this is often not the case in nature, especially for spatially close spins. In this paper, we experimentally and theoretically study the decay of such multi-qubit states under the action of a correlated spin bath.
In particular, we are interested in investigating the decay rate dependence on the correlations in a multi-qubit spin state.
We leverage the low dimensionality of the system studied -- the linear coupling geometry provided by
nuclear spins in apatite crystals~\cite{Cappellaro07a, Kaur12} --  to gain insight into both the many-body states created by the coherent Hamiltonian dynamics and their subsequent decay. We present a simple analytical model that captures the essential features of the multi-qubit decays,  and compares well with the experimental data. These results will be helpful in paving the way for the future design of schemes to mitigate the decay.

Specifically, in a linear spin system -- calcium Fluorapatite (FAp)~\cite{Teshima11} -- we consider multi-spin states created by the double quantum Hamiltonian ${\cal H}_{\mathrm{DQ}}$, which has been widely studied in quantum transport~\cite{Cappellaro07l}.
We analyze their decay under the natural dipolar Hamiltonian ${\cal H}_{\mathrm{dip}}$.
Similar decay dynamics were studied in more complex 3D spin systems~\cite{Krojanski04,Cho06,Lovric07,Sanchez07} and extended to  the study of localization phenomena \cite{Alvarez10c,Alvarez10d}.
However, in contrast to 3D systems, the highly restricted coupling topology in our experiments leads to analytically tractable solutions for the Hamiltonian evolution and the decay rates.
It is thus possible to have a much better characterization of the multi-spin states we can create than it was possible in more complex 3D systems.  As a result  we can study how the decay rate changes with the state characteristics, such as long-range correlations in extended spin clusters and their degree of coherence.  Focusing on 1D systems is also important in light of recent work in creating low-dimensional systems in ion-traps~\cite{Cetina13}, or via Hamiltonian engineering~\cite{Ajoy13l} in crystals.

The paper is organized as follows. We describe the experimental system and methods in Section~\ref{sec:Experiment} and we present the experimental results. In Section~\ref{sec:Theory} we introduce an analytical model (extended details can be found in Appendix~\ref{sec:Appendix}) to interpret the experimental results. This leads us to a better insight into the decay rate dependence on the state characteristics that we further explore in Section~\ref{sec:States} by experimentally studying a diverse set of states.

\section{Experimental Methods and Results}\label{sec:Experiment}
\subsection{The Spin System}
 The system of interest are $^{19}$F nuclear spins in a single crystal of fluorapatite (FAp) [Ca$_5$(PO$_4$)$_3$F]. We use Nuclear Magnetic Resonance (NMR) techniques to study the spin dynamics  at  room temperature  in a 300~MHz Bruker Avance Spectrometer ($B_0$=7T) with a probe tuned to 282.4~MHz.
In the FAp crystal, six parallel chains lie along the crystal $c$-axis, with a short intra-nuclear
spacing within a single chain,  $r_0=0.3442$nm, and a longer inter-chain separation of
$R=0.9367$nm.
The spins interact via the secular dipole-dipole Hamiltonian,
\begin{equation}
 {\cal H}_{\mathrm{dip}}=\sum_{ij}b_{ij}[2\sz^i\sz^j-(\sx^i\sx^j+\sy^i\sy^j)]
\label{eq:Hamiltonian}\end{equation}
where $\sigma_\alpha^i$ are the usual Pauli matrices   and
 the couplings are
 $$b_{ij}=\frac{\mu_0}{16\pi}\frac{\gamma^2 \hbar}{r_{ij}^3}(1-3\cos^2\theta_{ij}),$$
  with $\mu_0$  the standard magnetic constant,
$\gamma$ the $^{19}$F gyromagnetic ratio, $r_{ij}$ the distance between nucleus $i$
and $j$ and $\theta_{ij}$ the angle between $\vec r_{ij}$ and the
$z$-axis.
When the crystal is aligned with the magnetic field, as in our experiments, the nearest-neighbor inter-chain dipolar coupling
is about $40$ times weaker than the in-chain  coupling. Thus, for short evolution times, couplings across different chains can be neglected and the
system can be considered as a collection of one-dimensional spin chains~\cite{Cho93,Zhang09,Cappellaro07,RufeilFiori09,Kaur12}.

%%%%%%%%%%%%%%%%%%%%%%%%%%%%%%%%%%%%%%%%%%%%%%%%%%%%%%%%%%%%%%%%%%%%%%%%%%%%%%%
\begin{figure}[b]
 \centering
\includegraphics[width=0.45\textwidth]{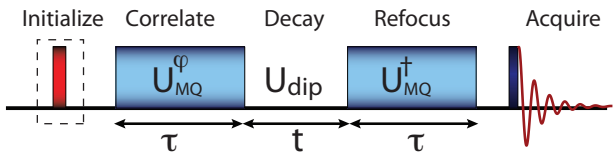}
\caption{Experimental scheme. The system is first prepared in an initial state of interest, for example the thermal equilibrium  state or $\sum\sx$ using a $\pi/2$-pulse (red bar).
Evolution via the DQ Hamiltonian ${\cal H}_{\mathrm{DQ}}$ (obtained by a multi-pulse sequence, blue  rectangles) creates spin correlations  during the time $\tau$. A phase shift $\phi$ of the propagator encodes information about the multiple quantum coherence intensities created.
The state undergoes decay under the dipolar interaction ${\cal H}_{\mathrm{dip}}$
during the time $t$. The correlated state is refocused by the inverse propagator $U_{\mathrm{DQ}}^\dag$ before a $\pi/2$ pulse is used to detect the spin free-induction decay.
}
\label{fig:sequence}
\end{figure}
%%%%%%%%%%%%%%%%%%%%%%%%%%%%%%%%%%%%%%%%%%%%%%%%%%%%%%%%%%%%%%%%%%%%%%%%%%%%%%%
\subsection{Experimental Protocol}
The experimental scheme  is shown in Fig.~\ref{fig:sequence}. The system is first prepared in a suitable initial state $\rho_i$ starting from its equilibrium thermal state. Evolution under a propagator $U_{\mathrm{MQ}}$ for a time $\tau$ creates a complex,  multiple-quantum coherence state~\cite{Yen83}. The system is then let evolve freely for a time $t$, during which the coherences decay mainly under the effects of the dipolar Hamiltonian. In order to observe this decay, we first refocus the remaining coherences with a propagator $U_{\mathrm{MQ}}^\dag$ before measuring the spin magnetization via the usual free induction decay.

The equilibrium state is the Zeeman thermal state, $\rho_{th}(0) \propto\exp(-\varepsilon \Sigma_z)\approx \Id - \varepsilon \Sigma_z$, where $ \Sigma_z= \sum_{j}\sz^j $ and $ \varepsilon = \gamma B_{0}/k_{B}T\ll1$ (with $k_B$ the Boltzmann constant and $T$ the temperature).
We focus on the evolution and decay of the deviation from identity of this state, i.e. $\delta\rho_{th}\sim\Sigma_z$, and of other initial states that can be created from $\delta\rho_{th}$ with appropriate manipulation (see Section~\ref{sec:States}). Indeed, the identity does not evolve and does not contribute to the signal.

Starting from the prepared initial state $\rho_i$, we create spin correlations by evolution under the double quantum (DQ) Hamiltonian
\begin{equation}
{\cal
H}_{\mathrm{DQ}} = \sum_{ij}b_{ij}(\sx^i\sx^j-\sy^i\sy^j),
\label{eq:dqham}\end{equation}
which is known to generate quantum coherences among the spins~\cite{Yen83}.
 The primitive pulse cycle is given by,
 $\mathrm{P}_2=\frac{\delta t}{2}\ \mathrm{-}\ \frac\pi2|_x\ -\ \delta t'\ -\ \frac\pi2|_x -\ \frac{\delta t}{2}$,
where $\delta t'= 2\delta t + w$, $\delta t$ is the delay between pulses and $w$ is the width the $\pi/2$ pulse. We used the symmetrized 8-pulse variant of this basic sequence~\cite{Yen83,Ramanathan03},
$\mathrm{P}_{8}=\mathrm{P}_2\cdot\overline{\mathrm{P}_2}\cdot \overline{\mathrm{P}_2}\cdot {\mathrm{P}_2}$
(where $\overline{\mathrm{P}_2}$ is the time-reversed version of
${\mathrm{P}_2}$),  which simulates ${\cal H}_{\mathrm{DQ}}$ to second order in the
Magnus expansion~\cite{Haeberlen76}.
In the experiments, the length of the $\pi$/2 pulse  was $w=1.01\mu$s and the evolution time was incremented by varying the inter-pulse delay from $\delta t=1\mu$s to $6.2\mu$s and the number of loops was increased from 1 to 12 (varying both parameters enabled exploring a wide range of evolution times up to 1ms).  A recycle delay of 5s was used to re-equilibrate the system.

\begin{figure*}[t]
   \centering\includegraphics[width=0.4\textwidth]{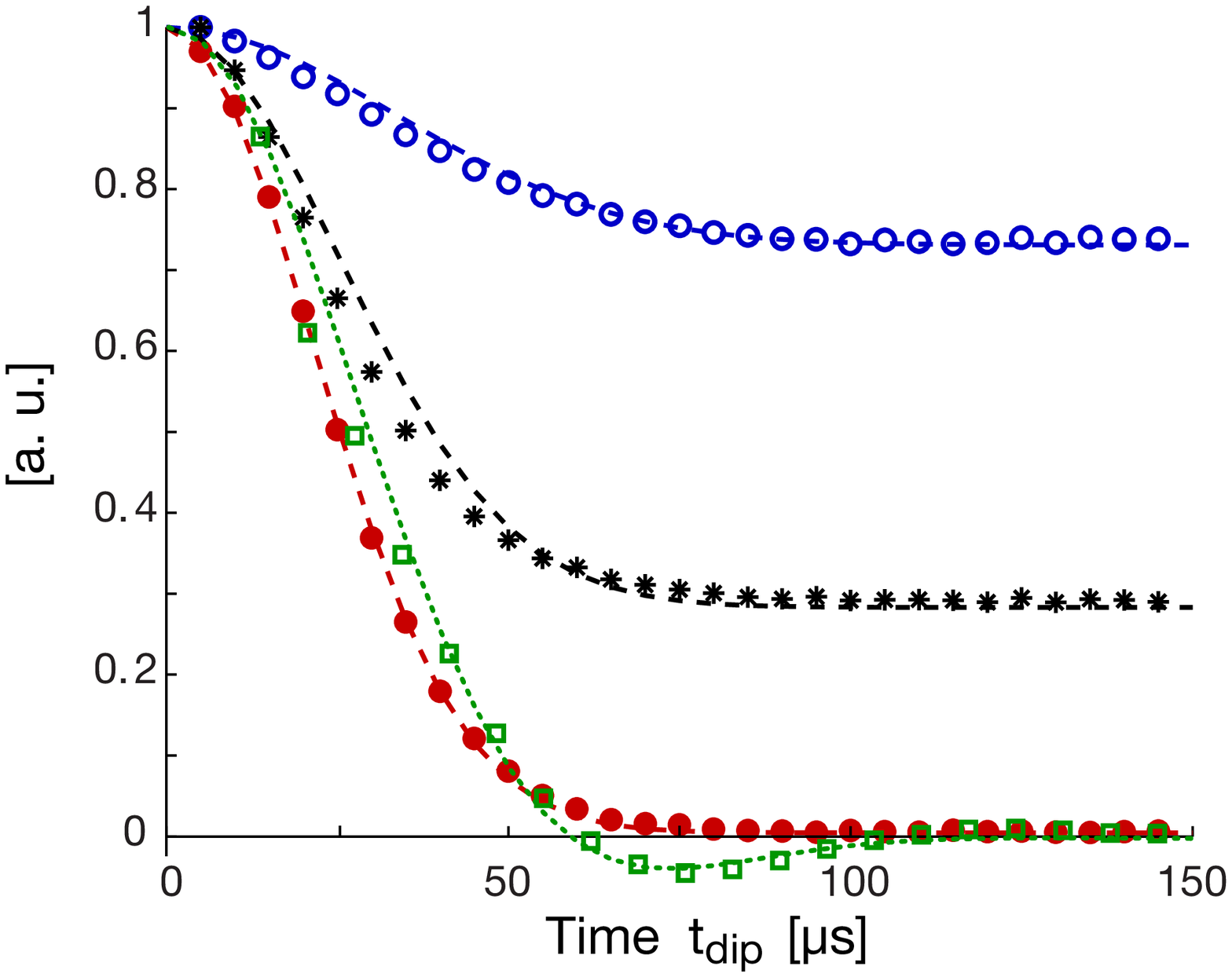}\qquad\qquad
    \centering\includegraphics[width=0.4\textwidth]{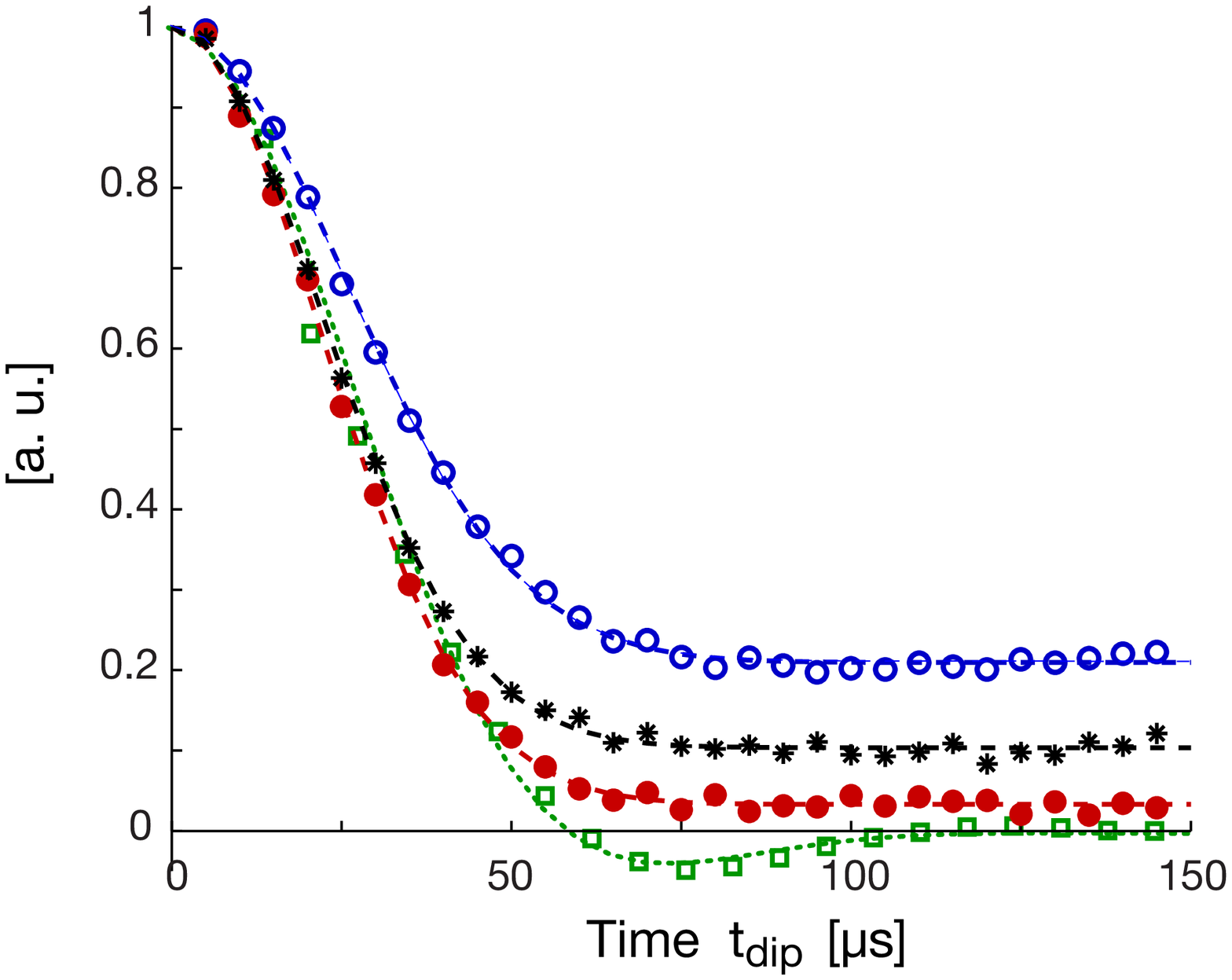}
   \caption{Normalized decay of quantum coherences created from the thermal initial state by evolving for a time $\tau$ under the $\mathcal{H}_{DQ}$ Hamiltonian. (ZQ (0Q) blue open circles, DQ (2Q) red dots, Total signal black stars).  Data points are the normalized signal intensities for evolution times $\tau=48\mu$s (left) and $\tau=589\mu$s (right). The dashed lines are fitting using the function in Eq.~(\ref{eq:Gmodel}).
For comparison we plot the free induction decay (green squares) that we fit with the function~\cite{Abragam82Book}
$A\left[(1-C)\mathrm{sinc}(m_2t)\,e^{-m_1t^2/2}+C\right]$,
with the second moment given by $M=m_1+m_2^2/3$.}
  \label{fig:decay}
 \end{figure*}

The  density operator created by evolution under the DQ Hamiltonian can be decomposed into its multiple quantum coherences (MQC) components, $\rho(\tau)=U_{\mathrm{MQ}}(\tau)\rho(0)U^\dag_{\mathrm{MQ}}(\tau) =
\sum_m\rho^{(m)}$, where a multiple quantum term of order $m$,  $\rho^{(m)}$, acquires a phase $m\phi$ under a collective $\Sigma_z$ rotation by an angle $\phi$.
The correlated spin states created under ${\cal H}_{\mathrm{DQ}}$ evolution contain in general  all \textit{even} $M$ coherence orders.
However, since standard NMR techniques measure only single-quantum
coherences (SQC), in order to  probe the higher spin coherences it is necessary to indirectly
encode their signatures into SQCs which can be measured
inductively~\cite{Baum85}. This is achieved by labeling each coherence order with a different phase $\phi$ by means of collective rotations $U_{\phi} = \exp(-i\phi\Sigma_{z}/2)$  about the z axis, effectively creating the phase shifted DQ Hamiltonian, ${\cal H}_{\mathrm{DQ}}^{\phi} = U_{\phi}{\cal H}_{\mathrm{DQ}}U_{\phi}^{\dag}$.
Finally, MQC are refocused back to
single-spin single-quantum terms and free induction decay is measured. Each measurement is repeated while incrementing $\phi$ from 0 to 2$\pi$ in steps of  $\delta\phi
= 2\pi/2K$ where $K$ is the highest order of MQC we wish to encode. If $\delta\rho_i$
is the initial density matrix,  the final density matrix $\delta\rho_f$ is given by:
\begin{equation}
\delta\rho_f(t,\tau)\!=\!
U_{\mathrm{MQ}}^\dag\!(\tau)U_{\mathrm{dip}}\!(t)
 U_{\mathrm{MQ}}^\phi\!(\tau) \delta\rho_i U_{\mathrm{MQ}}^{\phi\dag}\!(\tau)
U_{\mathrm{dip}}^\dag\!(t) U_{\mathrm{MQ}}\!(\tau),
\end{equation}
where $U_{\mathrm{MQ}}(\tau) = \exp(-i {\cal H}_{\mathrm{DQ}}\tau)$.

Since often the observable $\delta\rho_o$ is proportional to the initial state (as it is the case for the thermal equilibrium state and the total magnetization along the z-axis) we can write the measured signal as a correlation $S(t,\tau)=\Tr{\delta\rho(t,\tau)\delta\rho_o(\tau)}$, between the state prepared by the DQ evolution, $\delta\rho_o(\tau)=U_{\mathrm{MQ}}(\tau) \delta\rho_o U^\dag_{\mathrm{MQ}}(\tau)$ and the same state after decay under the dipolar evolution, $\delta\rho(t,\tau)=U_{\mathrm{dip}}\delta\rho_o(\tau) U_{\mathrm{dip}}^\dag$.
The signal intensities of various coherence orders are given by the Fourier Transform with respect to the phase $\phi$:
\begin{equation}
I^{(m)}(t,\tau)\!=\!\Tr{\delta\rho^{(m)}_o(\tau)\delta\rho^{(m)}(t,\tau)}\!=\!\sum_{k=1}^{K} S^k(t,\tau) e^{-ikm\delta\phi},
\label{eq:mqcint}
\end{equation}
where  $S^k(t,\tau)=\Tr{\delta\rho_f^k(t,\tau)\delta\rho_o}$ is the signal acquired in the $k$th measurement when setting $\phi=\pi k/K$.

%%%%%%%%%%%%%%%%%%%%%%%%%%%%%%%%%%%%%%%%%%%%%%%%%%%%%%%%%%%%%%%%%%%%%%%%%%%%%%%
\subsection{Results and Data Analysis}

We first studied the decay of MQC intensities created under ${\cal H}_{\mathrm{DQ}}$ starting from an initial thermal state ($\delta\rho_{th}\sim\Sigma_z$). In a quasi-1D system such as FAp, it is known that the DQ Hamiltonian excites mainly zero- and double-quantum coherences~\cite{Cho93,Feldman96}. The decay of the total signal, $S(t,\tau)=\Tr{\rho_f(t,\tau)\rho_i}$, and of each coherence intensity is shown in Fig.~\ref{fig:decay} as a function of dipolar decay time ($t$) for two exemplary MQC excitation times, $\tau=48\mu$s and $589\mu$s.
The decoherence dynamics was studied by repeating the experimental scheme described above while varying the DQ evolution time ($\tau$) from $36\mu$s to $925\mu$s and the decay time ($t$) from $0$ to $145\mu$s (which is on the order of the free induction decay time). 

\begin{figure*}[t]
   \centering\includegraphics[width=0.4\textwidth]{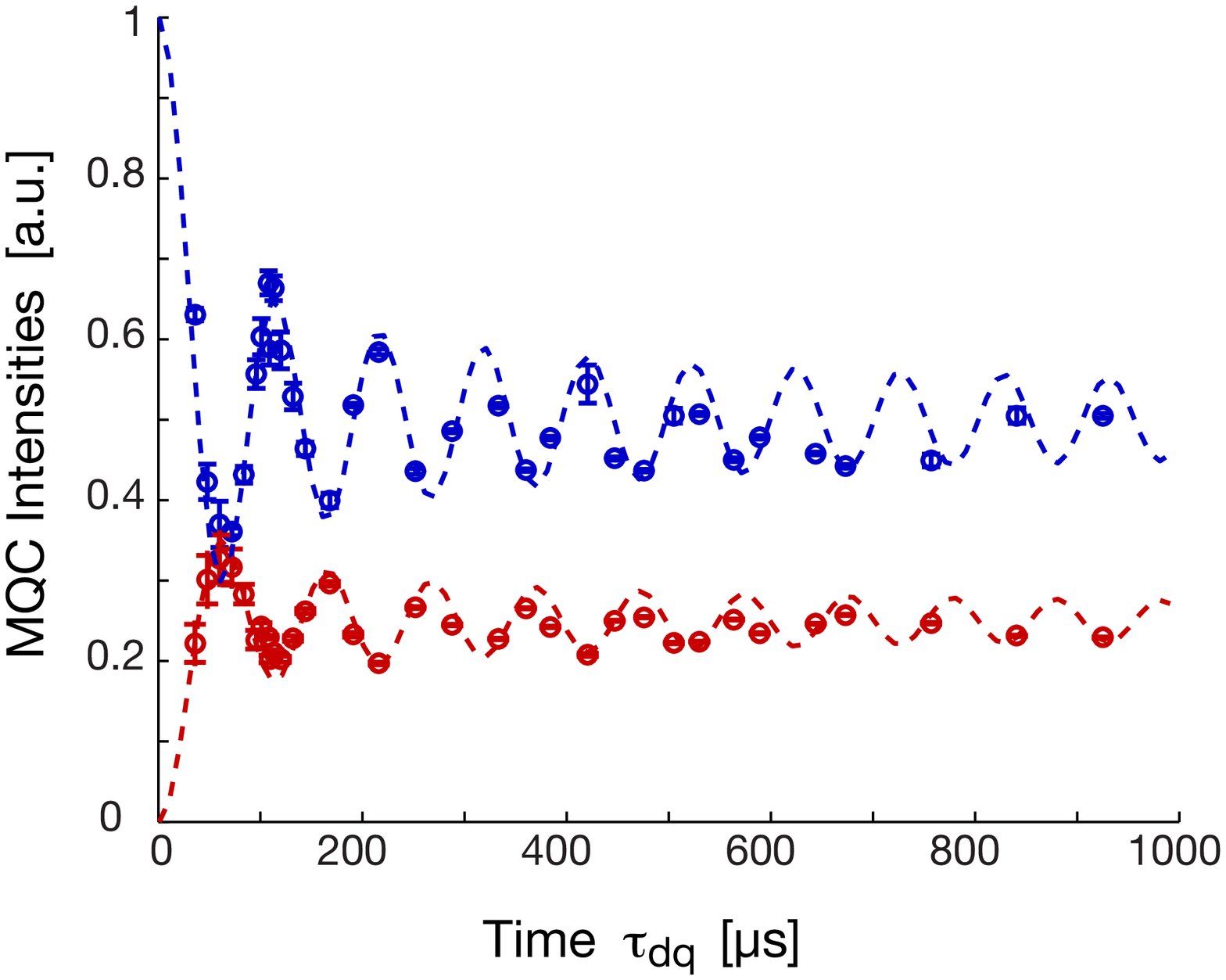}\qquad\qquad
      \includegraphics[width=0.4\textwidth]{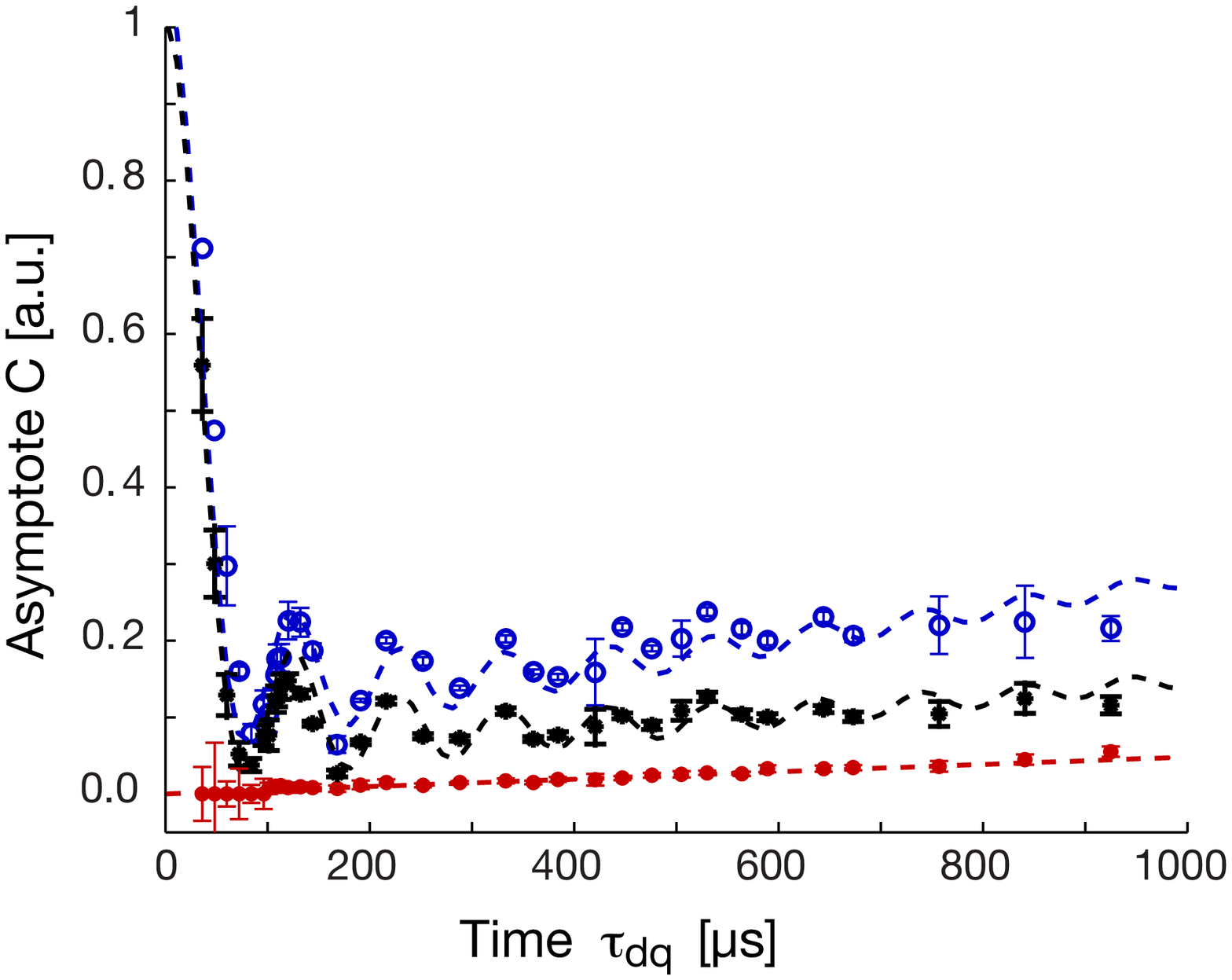}
   \caption{Left: Evolution of multiple quantum coherence intensities  ($A_{\mathrm{ZQ}}$ blue, $A_{\mathrm{DQ}}$ red) starting from Zeeman initial state.  The MQC intensities are obtained from the amplitude  parameter $A(\tau)$ when fitting the decay curves  (error bars are also obtained from the fitting).  The MQC intensities are then fitted by analytical functions (blue and red dashed lines) obtained from the analytical model of the DQ Hamiltonian with NN couplings (Eq. \ref{eq:I0}
-\ref{eq:I2}), yielding a dipolar coupling 7.7 $\times10^3$rad/s.   
Right:   Asymptotes of the experimental decay curves ($C_{\mathrm{ZQ}}$ blue,  $C_{\mathrm{DQ}}$ red $C_{\mathrm{total}}$ black). The curves were fitted by Eq.~(\ref{eq:cfit}) with an additional linear term in $\tau$, giving a  dipolar coupling of 7.676 $\times10^3$rad/s.}
 \label{fig:MQCC}
 \end{figure*}

We fitted the decay curves to Gaussian functions,
\begin{equation}
G(t,\tau)=A(\tau)\left([1-C(\tau)] e^{-M(\tau)t^2/2}+C(\tau)\right),
\label{eq:Gmodel}
\end{equation}
where $A$ (amplitude), $M$ (second moment) and $C$ (asymptote) are used as fitting parameters that vary with the DQ time $\tau$.
As shown by the decay curves in Fig.~\ref{fig:decay} and by the behavior of the fitting parameters in Fig.~\ref{fig:MQCC}-\ref{fig:Moment}, the system exhibits an interesting   dynamics as a function of the DQ-time evolution. 
This is in contrast to what was observed in 3D systems~\cite{Krojanski04,Cho06}, where the decay simply becomes monotonically faster as the DQ-time $\tau$ is increased.  The difference can be traced to the fact that the constrained coupling topology in 1D systems allow for a slower decay dominated by nearest-neighbor interactions, while in 3D systems the decay is more rapid and diffusion-like. We can compare for example the decay of CaF$_2$~\cite{Cho06} with FAp. While the minimum distance between nearest neighbors is quite similar ($r_{\text{CaF}_2}=0.27$nm versus $r_{\text{FAp}}=0.34$nm) the decay time is much faster, both for short  and especially for longer DQ-times, where it can be almost an order of magnitude faster.
In the following, we present a theoretical model of the observed behavior in 1D systems.

\section{Theoretical model and interpretation} \label{sec:Theory}
To gain insight into the decay behavior, we model both the dipolar Hamiltonian and the DQ Hamiltonian as 1D, nearest-neighbor (NN) interactions, neglecting the smaller contributions from long-range couplings in the chain and between chains. This approximation is justified in the short time limit~\cite{Zhang09}.

%%%%%%%%%%%%%%%%%%%%%%%%%%%%%%%%%%%%%%%%%%%%%%%%%%%%%%%%%%%%%%%%%%%%%%%%%%%%%%%
\subsection{Decay Amplitude}
The nearest-neighbor DQ Hamiltonian is known to be analytically solvable
in 1D by means of a Jordan-Wigner mapping~\cite{Jordan28} onto a system of free fermions.
 The  density operators for thermal 
 initial state evolving under the DQ Hamiltonian can be expressed in terms of canonical
fermionic  operators as~\cite{Ramanathan11,Doronin00}:
\begin{equation}
\begin{array}{ll}
\rho_{th}(\tau) =& \displaystyle\!\!\!\!\!\!\!\sum_{p-q\,\in\,\mathrm{even}}\!\!\!\!\! i^{-(p+q)}f_{pq}(2\tau)\left(c^\dag_pc_q+c_q^\dag c_p-\delta_{p,q}\right)\\& +i
\displaystyle\!\!\!\!\!\sum_{p-q\,\in\,\mathrm{odd}}\!\!\!\!\!i^{-(p+q)}f_{pq}(2\tau)(c^\dag_pc^\dag_q-c_qc_p),
\end{array}
\label{eq:rho}
\end{equation}
where the first term describes zero-quantum coherences, $\rho^{(0)}(\tau)$, and the second term double-quantum coherences, $\rho^{(2)}(\tau)$.
Here we defined the fermionic operators $c_p=\left(\prod_{k<p}\sz^k\right)\sm^p$ and
\begin{equation}
 f_{pq}(\tau) =\frac 2{n+1}\sum_k (-1)^p\sin(p\kappa)\sin(q\kappa)e^{-2ibt\cos\kappa},
\label{eq:Apq}
\end{equation}
 where $N$ is the chain length and $\kappa=\frac{\pi k}{N+1}$.

We are interested in following the decay of these states under the dipolar Hamiltonian. We thus consider the normalized
signal as a function of decay time $t$,
\begin{equation}
S(t,\tau) =\frac{
\Tr{\delta\rho_o(\tau)U_{\mathrm{dip}}(t)\delta\rho(\tau)U_{\mathrm{dip}}^\dag(t)}}{\Tr{\delta\rho_o(\tau)\delta\rho(\tau)}}
\end{equation}
We first note that the amplitude $A(\tau)$ in Eq.~(\ref{eq:Gmodel}) is given either by the total signal,
$A_S(\tau)=\Tr{\delta\rho_o(\tau)\delta\rho(\tau)}$ or by the MQC intensities, $A_{m\textsc{q}}(\tau)=\Tr{\delta\rho^{(m)}_o(\tau)\delta\rho^{(m)}(\tau)}$,  at $t=0$. This is shown in Fig.~(\ref{fig:MQCC}), where we plot $A_{\textsc{zq}}(\tau)$ and $A_{\mathrm{DQ}}(\tau)$. Since the total signal decays during the DQ evolution time $\tau$ (due to imperfection in the creation of the NN DQ Hamiltonian because of pulse errors, higher-order corrections in the Magnus expansion as well as errors due to long-range couplings~\cite{Zhang09}) we  normalize these amplitudes by the total signal amplitude $A_S(\tau)$. Upon this correction it is  possible to fit the amplitudes to well-known analytical solutions for the zero- ($I^{(0)}$) and double-quantum ($I^{(2)}$)  intensities~\cite{Feldman96} (see also  Appendix~\ref{sec:Appendix}).  
From these fits, considering an infinite chain, (Fig.~\ref{fig:MQCC}) we find a NN dipolar constant $b=7.7\times10^3$ rad/s. The value of the dipolar constant $b$ agrees very well with the one obtained from similar measurements done on a different FAp crystal~\cite{Zhang09,Kaur12} and also with the theoretical value $b=8.17\times10^3$rad/s obtained from the known structure of FAp.

\subsection{Long-time Asymptote}
%%%%%%%%%%%%%%%%%%%%%%%%%%%%%%%%%%%%%%%%%%%%%%%%%%%%%%%%%%%%%%%%%%%%%%%%%%%%%%%
%%%%%%%%%%%%%%%%%%%%%%%%%%%%%%%%%%%%%%%%%%%%%%%%%%%%%%%%%%%%%%%%%%%%%%%%%%%%%%%
\begin{figure*}[t]
 \centering
{\includegraphics[width=0.4\textwidth]{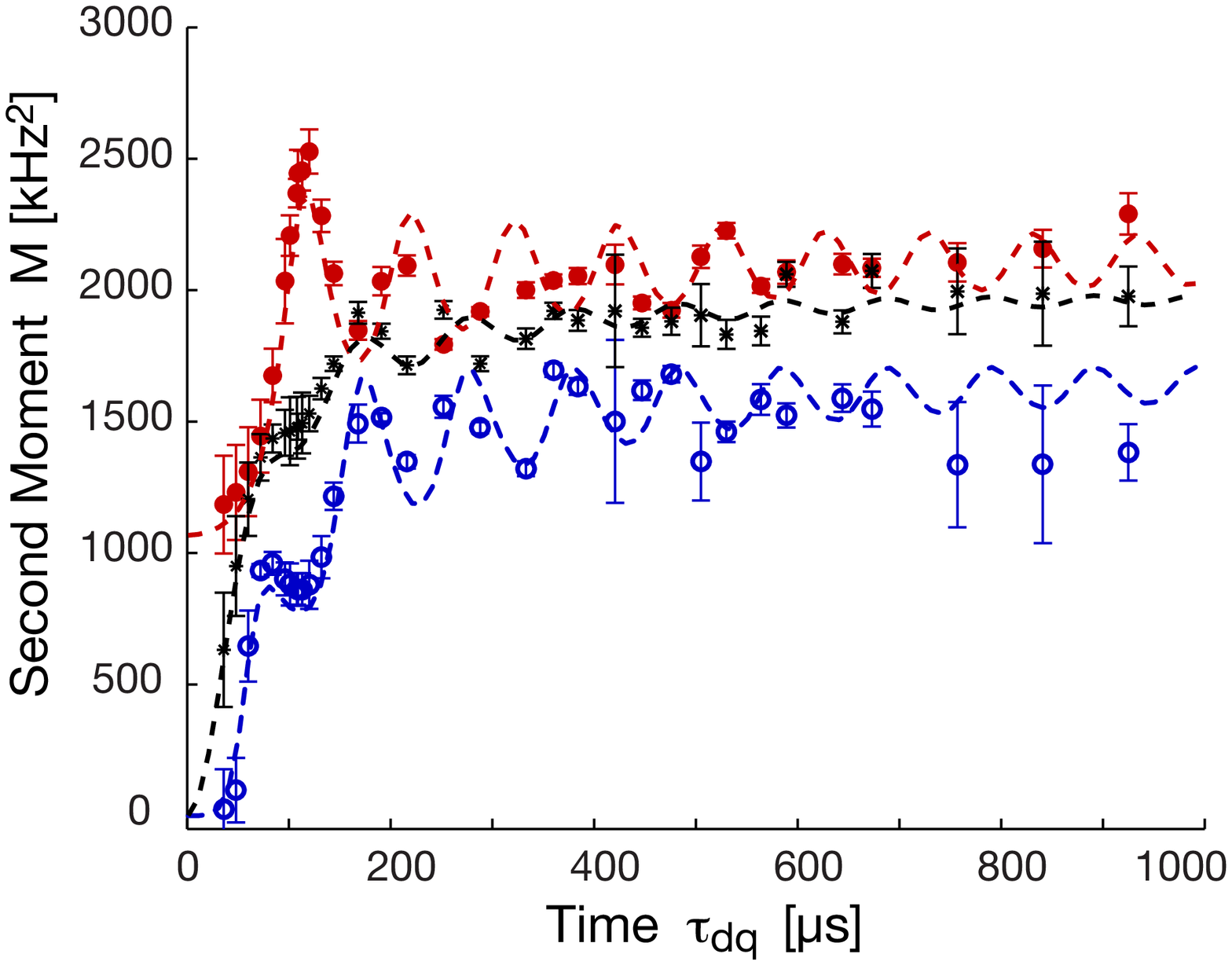}}\qquad\qquad
{\includegraphics[width=0.4\textwidth]{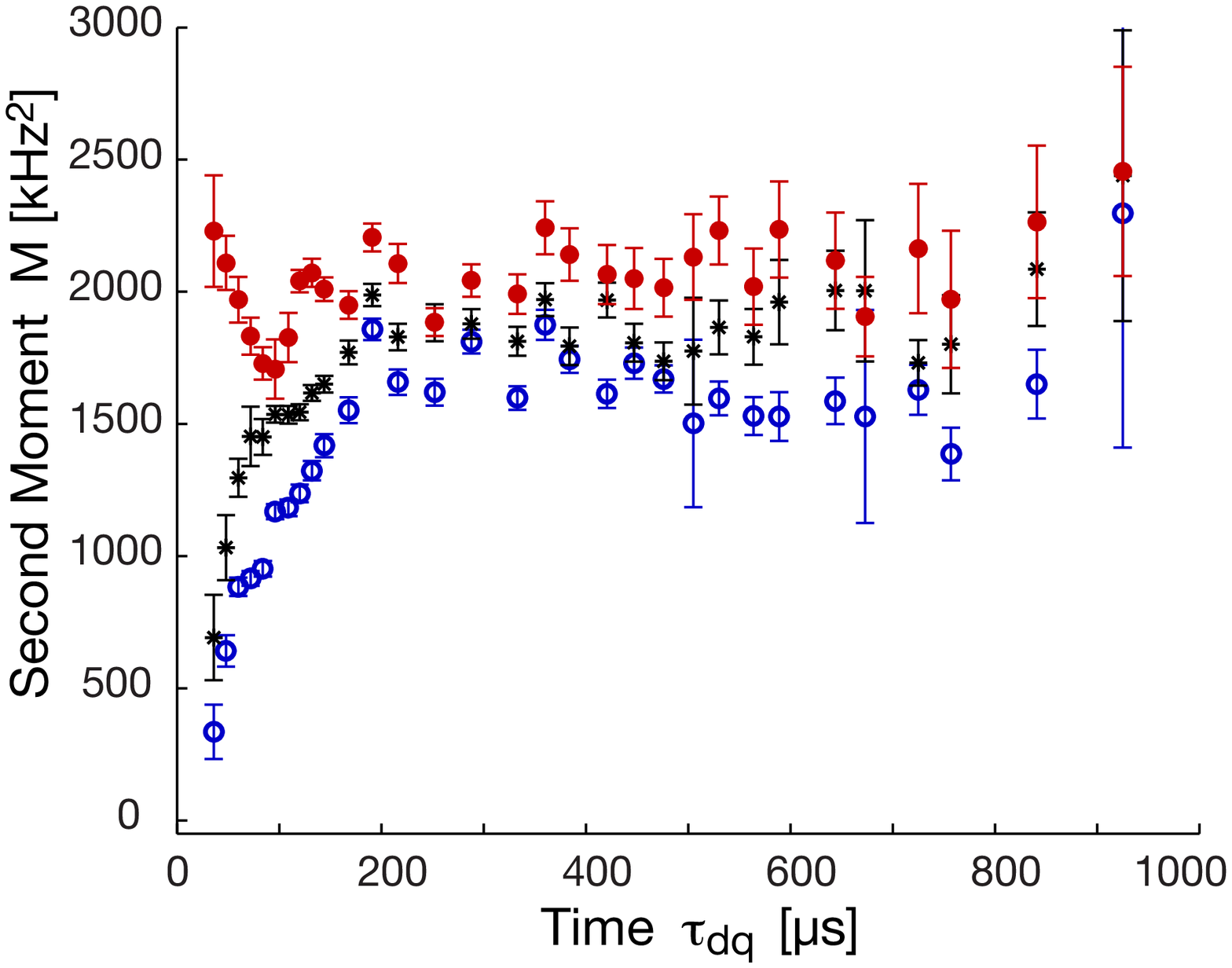}}
\caption{Decay rates of the different MQ components as a function of
double quantum evolution time $\tau$ for (left) the thermal and (right) end-polarized initial state. Points are experimental results obtained from the fitting of individual decay curves (such as in Fig.~\ref{fig:decay}) for each DQ-time $\tau$. Error bars are estimated from fitting of decay curves with ~\ref{eq:Gmodel}). The dashed
 lines are $\mu\, M(b,\tau)$, where $M(b,\tau)$ are the analytical curves in  Appendix~\ref{sec:Appendix}  for $N=100$ with $b=7.9\times10^3$rad/s. The total signal moment (black stars) was scaled by a factor $\mu=1.7$, the ZQ moment (blue circles) by a factor 1.8 and the DQ moment (red, dots) by 1.5.}
\label{fig:Moment}
\end{figure*}
%%%%%%%%%%%%%%%%%%%%%%%%%%%%%%%%%%%%%%%%%%%%%%%%%%%%%%%%%%%%%%%%%%%%%%%%%%%%%%%

The only terms that contribute to the asymptote $C$ are those that commute with the dipolar Hamiltonian, i.e. $C=\Tr{\rho^c_i(\tau)\rho_o^c(\tau)}/\Tr{\rho(\tau)\rho_o(\tau)}$,
such that $[{\cal H}_{\mathrm{dip}},\rho^c]=0$. In the NN approximation, $\rho^c$ contains only the population terms
(i.e. $p=q$ in Eq.~\ref{eq:rho}); hence we obtain
\begin{equation}
C(\tau)\!=\!\left( \frac1N\sum_p (-1)^p f_{pp}(2\tau)\right)^2\!\!=\!\left(\frac1N \sum_k \cos(4\tau\cos\kappa)\right)^2
\label{eq:cfit}
\end{equation}
As $\rho^c$ only contains zero-quantum terms, we expect a zero asymptote for the DQ intensities.
 We use the function $C(\tau)$ with an additional linear term in $\tau$ to fit the experimentally obtained asymptotes. This is shown in Fig.~(\ref{fig:MQCC}). The dipolar coupling (b) is used as a fitting parameter and the  value obtained from the fit is  $b=7.676\times10^3$rad/s, which
 agrees very well with the MQC fittings. The linear increase of the asymptote with time is due to errors in the implementations of the DQ evolution, as terms which are not zero- and double-quantum coherences appear as an increased population term when normalizing the signal.

\subsection{Decay Rate}
To determine the decay rate, we analyze the evolution by a short time expansion~\cite{Doronin04}, $\rho_i(t,\tau)\approx\rho_i(\tau)-it[{\cal H}_{\mathrm{dip}},\rho_i(\tau)]-\frac{t^2}2[{\cal H}_{\mathrm{dip}},[{\cal H}_{\mathrm{dip}},\rho_i(\tau)]]$, with a corresponding signal, $S(t,\tau)\propto\Tr{\rho_i(t,\tau)\rho_o(\tau)}$. We note  that the first order
term does not give any contribution to the signal~\cite{Ajoy12}, so we calculate the second moment 
\begin{equation}
\begin{array}{ll}
M = \displaystyle\frac{\Tr{[{\cal H}_{\mathrm{dip}},[{\cal H}_{\mathrm{dip}},\rho_i(\tau)]]\rho_o^{\dag}(\tau)}}{\Tr{\rho_i(\tau)\rho_o(\tau)^\dag}}\\
\displaystyle=\frac{\Tr{[\mathcal H_{\mathrm{dip}},\rho_i(\tau)][\rho_o(\tau),{\cal H}_{\mathrm{dip}}]^\dag}}{N2^N}
\end{array}
\label{eq:Moment}
\end{equation}
We can further calculate the contributions to the second moment arising from the zero- and double-quantum terms of the density operator
\begin{equation}
M^{(m)} = \frac{\Tr{[{\cal H}_{\mathrm{dip}},\rho^{(m)}(\tau)][\rho^{(m)}(\tau),{\cal H}_{\mathrm{dip}}]^\dag}}{\Tr{\rho^{(m)}(\tau)\rho^{(-m)}(\tau)}}
\label{eq:MomentMQC}
\end{equation}
These functions can be calculated analytically (see  Appendix~\ref{sec:Appendix}) thanks to the mapping to fermionic operators. 
We used these functions to analyze the second moments in Fig.~\ref{fig:Moment}, fitting the experimental momentum $M_{exp}$ to the analytical functions $\mu M(b)$.  From the fits we obtained $b=7.9\times10^3$rad/s and $\mu\approx1.7$. This indicates that while the variations of $M$ with the DQ evolution time $\tau$ are well in agreement with the analytical model and the expected dipolar coupling strength, the experimental second moment $M_{exp}$ is larger than expected from this model. This indicates that other  mechanisms contribute to the decay, including longer range couplings and control errors in the state preparation.

Some features of the second moments are worth pointing out. At small DQ times, the decay rate of the ZQ intensities (and of the signal) goes to zero, as indeed the initial state is an equilibrium state that commutes with the dipolar Hamiltonian. Instead,  $M^{(2)}$ has a finite asymptote, $b^2/12$, for $\tau\to0$ (see Eq.~\ref{eq:Mzzdq},\ref{eq:Mxxdq}): mathematically, this is because both the commutator and the DQ intensities go to zero with $\tau$; physically, this means that as soon as some DQ term is created in the state, it will decay with a finite rate under the dipolar Hamiltonian.

The second moments of both MQC intensities then oscillate in time with $\tau$, with the DQ moment always being larger than the ZQ one, $M^{(2)}>M^{(0)}$. This can be understood by their different behavior under the dipolar Hamiltonian. Consider the ZQ state, $\rho_{th}^{(0)}(\tau)$ (first line in Eq.~\ref{eq:rho}) and the ``flip-flop'' term of the dipolar Hamiltonian, ${\cal H}_{xx}=b\sum_j(c^\dag_jc_{j+1}+c^\dag_{j+1}c_j)$.  If we had considered periodic boundary conditions (instead of an open chain), these two operators would have commuted. Thus we expect their contribution to the second moment to be small and decreasing with the chain length $N$. In contrast, the contribution of ${\cal H}_{xx}$ to the second moment of $\rho^{(2)}(\tau)$ is on the same order as the contribution from ${\cal H}_{zz}=b\sum_j\sz^j\sz^{j+1}$, thus yielding an overall faster decay rate. We can understand this behavior more intuitively. Here we defined quantum coherence with respect to the total magnetization along the z-axis~\cite{Cho05}, $\Sigma_z$, which also sets the quantization axis of the system. Indeed, we only retained the part of the dipolar interaction that commutes with $\Sigma_z$. Thus we expect terms in the system state that commutes with $\Sigma_z$ (such as ZQ terms) to decay more slowly than terms that do not.

It is interesting to note that  both the ZQ and DQ second moments are higher when the corresponding MQC intensity is smaller. By analyzing the state in Eq.~\ref{eq:rho}, we can see that times $\tau$ of local maxima for $I^{(0)}(\tau)$ correspond to maxima for the function $f_{pp}(2\tau)$, which indicates that the state is localized around single-spin polarization states, $\sz^j$, while at other times the ZQ intensities have stronger contributions from many-spin correlations (three and higher). Correspondingly, local maxima for $I^{(2)}(\tau)$ correspond to higher contributions from localized two-spin double quantum states in $\rho(\tau)$, $\sp^j\sm^{j+1}$.
We thus found that the decay rates depends not only on the coherence order of the state,  but also on whether the state is localized or contains a larger number of spin correlations. While these two effects could not  be distinguished clearly in 3D experiments, as it was not possible to determine the precise states created,  the one-dimensional case provides further insights into the decay dynamics. We can further explore these differences by studying the decay of other correlated states.

\section{Comparison of initial states}\label{sec:States}

In order to study the dependence of the decay rate on the coherence order and the number of correlated spins, we evolved different initial states under the DQ Hamiltonian.
Besides the thermal state, we considered two other initial states: the ``end-polarized'' state $\delta\rho_{\mathrm{end}}\propto(\sz^1+\sz^N)$~\cite{Cappellaro07a,Kaur12} and a state rotated in the transverse plane, $\Sigma_x=\sum_j\sx^j$. We prepared the first state by the two-pulse scheme introduced in Ref.~\cite{Cappellaro07a}, while the second state can be prepared by a simple collective rotation of the spins.

The end-polarized state exhibits a transport-like dynamics under the DQ Hamiltonian~\cite{Cappellaro07l}; although the transport is dispersive~\cite{Cappellaro11,Ramanathan11}, we expect the state to remain fairly localized at short times and thus to show similar decay behavior as the thermal state. In contrast, the initial state $\delta\rho_x\sim\Sigma_x$ quickly evolves into many-spin correlations. Indeed, if we consider for example evolution of the first spin in the chain $\delta\rho_{x_1}\sim\sx^1$, we obtain:
\begin{equation}\begin{array}{ll}
\delta\rho_{x_1}\!(\tau)=&\displaystyle\sum_{p\,\in\,\mathrm{odd}}\!\!\!\mathrm{Im}[i^{-p}f_{1p}(\tau)](c_p^\dag+ c_p)\\&- i\displaystyle\sum_{p\,\in\,\mathrm{even}}\!\!\!\mathrm{Im}[i^{-p}f_{1p}(\tau)](c_p^\dag- c_p)\end{array}
\label{eq:sigmaX}
\end{equation}
Here we note that the fermionic operators $c_p$ represent highly delocalized states, since we have e.g. $c_p^\dag+ c_p=\sz^1\cdots\sz^{p-1}\sx^p$. Similar expressions can be found for the evolution of the other spins in the chain and thus for $\delta\rho_x(\tau)$. While this state presents large spin correlations, its coherence number is still quite low, with mostly one- and three-quantum coherences~\cite{Ramanathan11}.
In order to investigate the decay of larger coherence orders, we thus rotate the state $\delta\rho_x(\tau)$ with a $\pi/2$-pulse around the y-axis before letting it evolve freely. The resulting state, $\delta\rho_{xx}$, contains all even quantum coherence orders with a binomial distribution and it is thus more similar to the states that can be obtained in 3D systems~\cite{Baum85}.

\begin{figure}[t]
 \centering
\includegraphics[scale=0.33]{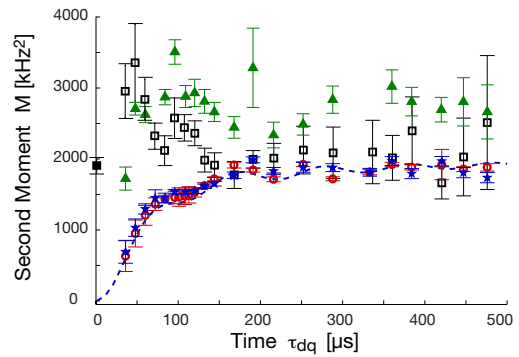}
\caption{Decay rates of the total signal as a function of evolution time $\tau$ for
different initial states: thermal state $\delta\rho_{th}$ (blue stars), end-polarized state $\delta\rho_{\mathrm{end}}$ (red circles); transverse polarization, $\delta\rho_x$ (black squares) and the state presenting a broader distribution of MQC, $\delta\rho_{xx}$ (green triangles). The black filled square at $\tau=0$ is the second moment of the free-induction decay.
As the $\delta\rho_x(\tau)$ states present oscillations in the decay (similar to the free-induction decay, see Fig.~\ref{fig:decay}) we fitted their decay to the function~\cite{Abragam82Book}
$A\left[(1-C)\mathrm{sinc}(m_2t)\,e^{-m_1t^2/2}+C\right]$,
with the second moment given by $M=m_1+m_2^2/3$.
The dashed line is the analytical model for the thermal state second moment, as in Fig.~\ref{fig:Moment}.}
\label{fig:states}
\end{figure}

We compare the decay rates (second moment) of these four different states in Fig.~(\ref{fig:states}). When the evolution under the DQ-Hamiltonian is short, the decay rate of $\delta\rho_{\mathrm{th}}$ and $\delta\rho_{\mathrm{end}}$ is small, as expected. Although $\delta\rho_{xx}$ has a small decay rate at short times, since it has mainly contributions from ZQ coherences, the second moment increases quickly with $\tau$. In comparison, $\delta\rho_{x}$ has a fast decay even when it has not evolved under the DQ-Hamiltonian (indeed for $\tau=0$ we recover the second moment of the free-induction decay). At larger $\tau$ the second moment  still remains  slightly larger than the second moments of $\delta\rho_{\mathrm{th}}$ and $\delta\rho_{\mathrm{end}}$. Indeed at larger $\tau$ all the states becomes fairly delocalized because of the dispersive character of the equal-coupling DQ Hamiltonian~\cite{Cappellaro11}. Thus these different states highlight different behaviors of multi-spin correlated states, which depend on the number of spins in the correlated state (with faster decay for larger spin correlation number) and separately on the coherence number.
%%%%%%%%%%%%%%%%%%%%%%%%%%%%%%%%%%%%%%%%%%%%%%%%%%%%%%%%%%%%%%%%%%%%%%%%%%%%%%%
%%%%%%%%%%%%%%%%%%%%%%%%%%%%%%%%%%%%%%%%%%%%%%%%%%%%%%%%%%%%%%%%%%%%%%%%%%%%%%%

\section{Conclusions}
In this paper we investigated the dependence of decoherence rate on the state characteristics of a many-spin system. Since the decay process is non-Markovian, but it is due to a highly correlated spin bath, we found a very rich dynamics, where decoherence rates (quantified by the second moment of the the decay) depend in a non-trivial way on the degree of localization of the state as well as on its coherence with respect to the quantization basis.
In particular we found that large spin clusters, with correlations established among many spins, decay faster under a correlated bath, even if their coherence order is not very large. This is in contrast to the decay under simple dephasing, where the coherence order (and for pure states, the entanglement) is critical in determining the decay rate~\cite{Huelga97}.  While it was not possible to separate the coherence order and the number of correlated spins in the dynamics of 3D spin systems (as they grow at the same time), here we were able to get  more insight by using spin chains and exploring different initial states. In addition we found that restricting the dynamics in one dimension slows down the decay, which could be beneficial to create larger coherent quantum states.
%=========================================================================
\begin{acknowledgements}
This work was partially funded by NSF under grant DMG-1005926 and by AFOSR YIP.  
\end{acknowledgements}

%=========================================================================

\appendix
\section{Calculation of the second moment}\label{sec:Appendix}
Here we provide explicit expressions for the second moment of the  decay under the dipolar Hamiltonian of the thermal state evolved under the DQ Hamiltonian for a time $\tau$. Using the second moment to estimate the decay rate is justified by a  a short time expansion of the signal,
\begin{eqnarray}
&&S(t,\tau)\approx S(\tau)\!\left(1-M\frac{t^2}2\right)\\&&\quad=\Tr{\delta\rho(\tau)^2}\!\left(1-\frac{\Tr{[{\cal H}_{\mathrm{dip}},\rho(\tau)][\rho(\tau),{\cal H}_{\mathrm{dip}}]^\dag}}{\Tr{\delta\rho(\tau)^2}}\frac{t^2}2\right)\nonumber
\end{eqnarray}
We can  calculate the contributions to the second moment arising from the zero- and double-quantum terms of the density operator
\begin{equation}
M^{(m)} = \frac{\Tr{[{\cal H}_{\mathrm{dip}},\rho^{(m)}(\tau)][\rho^{(m)}(\tau),{\cal H}_{\mathrm{dip}}]^\dag}}{\Tr{\rho^{(m)}(\tau)\rho^{(-m)}(\tau)}},
\label{eq:MomentMQC}
\end{equation}

Further, writing the dipolar Hamiltonian as
 \begin{equation}\begin{array}{c}
\displaystyle{\cal
H}_{\mathrm{dip}}=2{\cal H}_{zz} -{\cal H}_{xx}, \\ \displaystyle{\cal H}_{zz}=\sum_{ij}b_{ij}\sz^i\sz^j,\quad{\cal H}_{xx}=\sum_{ij}b_{ij}(\sx^i\sx^j+\sy^i\sy^j)\end{array}
\label{eq:Hamiltonian}\end{equation}
we can also separate the contributions from the ${\cal H}_{zz}$ and ${\cal H}_{xx}$ parts of the dipolar Hamiltonian,
as they simply add up (there are no contributions from cross-terms). We can thus define each contribution as
\begin{equation}
M^{(n)}_{aa}=\frac{\tr{[{\cal H}_{aa},\rho^{(n)}(\tau)][{\cal H}_{aa},\rho^{(n)}(\tau)]^\dag}}{(N2^N)},
\label{eq:M2term}
\end{equation}
yielding the total second moment:
\begin{eqnarray}
&&M=M_{zz}+2M_{xx}\\&&
=\frac{I^{(0)}(M^{(0)}_{zz}+2M^{(0)}_{xx})+I^{(2)}(2M^{(2)}_{zz}+4M^{(2)}_{xx})}{N2^N},\nonumber
\end{eqnarray}
where $I^{(m)}$ are the MQC intenties~\cite{Feldman96}
\begin{equation}
I^{(0)}(\tau)=\frac1N \sum_k \cos(4\tau\cos\kappa)^2
\label{eq:I0}
\end{equation}
\begin{equation}
I^{(2)}(\tau)=\frac1{2N} \sum_k \sin(4\tau\cos\kappa)^2.
\label{eq:I2}
\end{equation}
\begin{figure*}[t]
   \centering
  \includegraphics[scale=0.33]{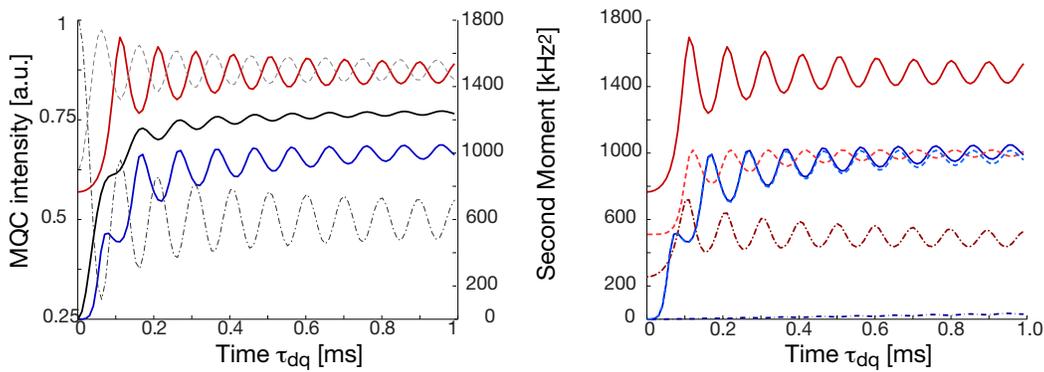}
\caption{Analytical solutions to second moments for $N=400$ spins. Left: second moment for the ZQ intensities (blue), DQ (red) and total signal (black).  In lighter gray: MQC intensities ($I^{(0)}$, ZQ, dash-dotted line, $I^{(2)}$, DQ, dashed line) highlighting the anti-correlation with the momentum oscillations.  For an easier visualization, we plot $I^{(2)}+ 0.625$.
Right: Contributions of the $\ham_{xx}$  (dash-dotted) and $\ham_{zz}$  (dashed)  Hamiltonians to the total (solid) second momenta for ZQ  (blue lines) and DQ (red lines) intensities.
We note that the second moment from $\ham_{xx}$ is almost zero for the ZQ intensities, the small increase at larger DQ times $\tau_{dq}$ is due to finite-length chain effects.}
  \label{fig:decayanal}
 \end{figure*}

\begin{widetext}
Using the state in Eq.~\ref{eq:rho} we find the contribution from $\ham_{zz}$:
\begin{equation}
M_{zz}=\frac{16}N\left[\sum_{p\neq q}|f_{p,q}(2\tau)|^2-\sum_{q=2}^N \left(|f_{q,q-1}(2\tau)|^2+ |f_{q,q-2}(2\tau)|^2+ |f_{1,q}(2\tau)|^2\right)\right],
\label{eq:Mazz}
\end{equation}
which is given by the sum of the ZQ and DQ contributions:
\begin{equation}
M_{zz}^{(0)}I^{(0)}=\frac{16}N\left[\sum_{\stackrel{p\neq q,}{p-q\in\,\mathrm{even}}}|f_{p,q}(2\tau)|^2-\sum_{q=2}^N |f_{q,q-2}(2\tau)|^2-\sum_{\stackrel{p\neq1,}{p\,\in\,\mathrm{odd}}}|f_{1,q}(2\tau)|^2\right]
\label{eq:Mzzzq}
\end{equation}
\begin{equation}
M_{zz}^{(2)}I^{(2)}=\frac{8}N\left[\sum_{p-q\in\,\mathrm{odd}}|f_{p,q}(2\tau)|^2-\sum_{q=1}^N |f_{q,q-1}(2\tau)|^2-\sum_{p\,\in\,\mathrm{even}}|f_{1,p}(2\tau)|^2\right]
\label{eq:Mzzdq}
\end{equation}
From the commutator with $\ham_{xx}$ we obtain
\begin{equation}
%\begin{array}{ll}
M_{xx}=\frac{4(N\!-\!1)}N\!-\!\frac{2}N\sum_{p,q}\left[f_{p+1,q}(2\tau)\!-\!f_{p-1,q}(2\tau)\right]\left[f_{p,q+1}(2\tau)\!-\!f_{p,q-1}(2\tau)\right],
\label{eq:Mxx}
\end{equation}

which can be further decomposed into the ZQ and DQ components:
\begin{equation}
M_{xx}^{(0)}I^{(0)}=M_{xx}(\tau)-I^{(2)}(\tau)M_{xx}^{(2)}(\tau)\end{equation}
\label{eq:Mxxzq}
\begin{equation}
M_{xx}^{(2)}I^{(2)}=
\frac{2}N\sum_{p-q\,\in\,\mathrm{odd}}|f_{p,q}(2\tau)-f_{q-1,p+1}(2\tau)|^2,
\label{eq:Mxxdq}
\end{equation}
\end{widetext}

\bibliographystyle{apsrev4-1}
\bibliography{/Users/pcappell/Documents/Work/Papers/Biblio}

\end{document}